\documentclass[submission]{eptcs}

\providecommand{\event}{PLACES 2020} \usepackage{breakurl}             \usepackage{subcaption}
 \usepackage{cite}
 \usepackage[inline]{enumitem}
\usepackage{booktabs}
\usepackage{amsmath,amssymb,amsfonts}
\usepackage{algorithm2e}
\usepackage{afterpage}
\usepackage{graphicx}
\usepackage{amsmath} \usepackage{tikz}
\usepackage{pifont}
\usetikzlibrary{arrows,automata,backgrounds,calc,positioning,shapes,fit,shapes.multipart,shadows}
\usepackage[framemethod=TikZ]{mdframed}
\usepackage{textcomp}
\usepackage{xcolor}
\usepackage{listings, multicol}
\usepackage{url}
\usepackage{golang}
\usepackage{upgreek} \usepackage[draft]{fixme} \usepackage{url}
\usepackage[]{hyperref}
\usepackage[numbers]{natbib}
\usepackage{wrapfig}
\newcounter{example}[section]
\newenvironment{example}[1][]{\refstepcounter{example}\par\medskip
  \noindent \textbf{Example~\theexample. #1} \rmfamily}{\medskip}

\usepackage[normalem]{ulem}
\makeatletter
\newcommand\reduline{\bgroup\markoverwith {\lower3.5\p@\hbox{\sixly \textcolor{red}{\char58}}}\ULon}\font\sixly=lasy6 \makeatother 

\newcommand{\lstCodeSize}{\normalsize}
\newcommand{\lstPrimitiveStyle}{\color{blue}\bfseries}

\newcommand{\lstNumberStyle}{\tiny\sffamily\color{gray}}

\lstdefinestyle{nonumber}{numbers=none,
  xleftmargin=1em,
  framexleftmargin=1em,
}

\newcommand{\bnfor}{\; \mid \;}
\newcommand{\m}[1]{{\texttt{\textbf{#1}}}}
\newcommand{\colorize}[1]{{\color{purple}{#1}}}
\newcommand{\promelacolor}[1]{{\color{blue}{#1}}}

\newcommand{\minigo}{\textit{MiniGo}}
\newcommand{\togology}{\textsc{gomela}}
\newcommand{\gomela}{\togology}
\newcommand{\cmark}{\ding{51}}\newcommand{\xmark}{\ding{55}}

\newcommand{\num}{n}
\newcommand{\expr}{e}
\newcommand{\stmt}{s}
\newcommand{\stmts}{\tilde s}

\newcommand{\chan}{ch}
\newcommand{\var}{v}
\newcommand{\chanorvar}{x}

\newcommand{\default}{\m{default} : \stmts}

\newcommand{\args}{\widetilde \chanexpr}
\newcommand{\chanexpr}{a}

\newcommand{\coloneqq}{:=}
\newcommand{\forstmts}{r}
\newcommand{\for}{\m{\text{for}} \ \var \texttt{:=} \expr_1 \m{;} \expr_2 \m{;} \forstmts \, \{\stmts_3\}}
\newcommand{\commclause}{c}
\newcommand{\commclausedef}{\m{case } \commstmt : {\stmts}}

\newcommand{\commstmt}{\alpha}

\newcommand{\prog}{p}

\newcommand{\ifff}{\m{if}\ \expr \ \m{then} \ \stmts_1}
\newcommand{\ifelse}{\ifff \ \m{else}\ \stmts_2 }
\newcommand{\id}{id}
\newcommand{\select}{\m{select}\{\widetilde{\commclause}\}  }
\newcommand{\makechan}{\chan \texttt{ := } \newch}
\newcommand{\newch}{\m{make}(\m{chan},\expr)}
\newcommand{\true}{\m{true}}
\newcommand{\false}{\m{false}}
\newcommand{\close}{\m{close}(\chan)} 

\newcommand{\funcalldef}{\id(\args)}
\newcommand{\gostmt}{\m{go } \funcalldef}

\newcommand{\fundecl}{d}

\newcommand{\fundecldef}{\m{func} \ \id (\tilde{\chanorvar}) \, \{ \stmts \}}

\newcommand{\chansfunc}{\texttt{chans}}

\newcommand{\chanParams}{chanParams}

\SetKwIF{If}{ElseIf}{Else}{\colorize{if}}{\colorize{then}}{\colorize{otherwise if}}{\colorize{otherwise}}{} 
\SetKwSwitch{Switch}{Case}{Other}{\colorize{switch}}{\colorize{:}}{\colorize{case}}{\colorize{otherwise}}{}{}
\SetKwFor{ForEach}{\colorize{foreach}}{\colorize{:}}{}
\SetKwProg{Fn}{\colorize{function}}{}{}
\SetKwFunction{Flatten}{flatten}
\SetKwFunction{isstruct}{is\_struct}
\SetKwFunction{ischan}{is\_channel}
\SetKwFunction{fields}{structFields}
\newcommand{\kwlet}{{\bf{\colorize{let}}}}
\newcommand{\kweq}{{\bf{\colorize{=}}}}
\newcommand{\kwin}{{\bf{\colorize{in}}}}
\newcommand{\append}{}

\newcommand{\node}{v}

\newcommand{\closedChan}{\texttt{chanMonitor}}

\newcommand{\equals}{\texttt{==}}

\newcommand{\upperBound}[1]{\uparrow #1}
\newcommand{\lowerBound}[1]{\downarrow #1}

\newcommand{\kbs}{\Delta}
\newcommand{\translateBlockStmt}{\texttt{TransStmts}}

\newcommand{\mapcontains}{\m{lookup}}
\newcommand{\spawns}{spawns}

\newcommand{\body}{\texttt{body}}

\newcommand{\bound}{b}

\newcommand{\domain}{\textit{dom}}

\newcommand{\smallpara}[1]{\smallskip
  \noindent
  \textbf{#1}\ }
 \fxsetup{theme=color,mode=multiuser}
\FXRegisterAuthor{JL}{aJL}{JL}
\FXRegisterAuthor{ND}{aND}{ND}
\def\BibTeX{{\rm B\kern-.05em{\sc i\kern-.025em b}\kern-.08em
    T\kern-.1667em\lower.7ex\hbox{E}\kern-.125emX}}

\newcommand{\gomelatitle}{Bounded verification of message-passing concurrency in Go using Promela and Spin}

\title{\gomelatitle}

\author{Nicolas Dilley
\institute{
University of Kent
} 
\and
Julien Lange
\institute{
University of Kent 
}
}

\def\titlerunning{\gomelatitle}
\def\authorrunning{N. Dilley \& J. Lange}
\begin{document} 
\maketitle
\begin{abstract}
  This paper describes a static verification framework for the
  message-passing fragment of the Go programming language.
Our framework extracts models that over-approximate the
  message-passing behaviour of a program.
These models, or behavioural types, are encoded in Promela, hence
  can be efficiently verified with Spin.
We improve on previous works by verifying programs that include
  communication-related parameters that are unknown at compile-time,
  i.e., programs that spawn a parameterised number of threads or that
  create channels with a parameterised capacity.
These programs are checked via a bounded verification approach
  with bounds provided by the user.
\end{abstract}
 \section{Introduction}

Go is an increasingly popular programming language that is known for
its lightweight threads (called \emph{goroutines}) and native support
for message-passing concurrency.
Go programmers are encouraged to coordinate threads by
exchanging messages over channels, rather than using shared memory
protected by mutexes~\cite{goproverb}.
In a recent empirical survey~\cite{NDilley19}, we have discovered that
more than 70\% of the most popular Go projects on GitHub use
message-passing primitives.
Additionally, Tu et al.~\cite{Tu19} showed that message-passing based
software is as liable to errors as other concurrent programming
techniques. They also showed that Go concurrency bugs are hard to
detect and have a long life time.
This is reflected in a recent survey amongst Go programmers reporting
that programmers often do not feel they are able to effectively repair
bugs related to Go's concurrency features~\cite{web:go-survey}.
Concretely, message-passing concurrency bugs in Go fall in two
categories: ($i$) blocking errors, where a goroutine is permanently
waiting for a matching send/receive action and ($ii$) channel errors,
where a goroutine attempts to close or send to a channel that is
already closed.

The Go ecosystem provides little support for users to detect
concurrency bugs.
Its type system only ensures that each channel instance carries a single
specified data type. While a \emph{run-time} global deadlock detector
is available, it is silently disabled by some libraries.
To help programmers produce correct concurrent software, several
authors have proposed techniques to verify Go programs both statically
(at
compile-time)~\cite{Ng15,Lange17,Lange18,Stadtmuller16,Midtgaard18}
and dynamically (at run-time)~\cite{Sulzmann17,Sulzmann18}.
One of the more mature techniques for statically verifying Go programs
is Godel~\cite{Lange18} which relies on the similarity of Go's
message-passing aspect to CCS~\cite{Milner84}.
Godel follows an approach based on behavioural types where Go programs
are over-approximated by CCS-like processes, which in turns are
model-checked, using mCRL2~\cite{CranenGKSVWW13} for safety and
liveness properties.
Because mCRL2 only deals with finite-state models, Godel has
one key limitation: it does not support programs that spawn new
threads in for-loops, e.g., the program in
Figure~\ref{fig:for-example} is not supported.
This restriction limits the applicability of Godel to real-world
code-bases. Indeed, 58\% of the Go projects we studied
in~\cite{NDilley19} feature thread-spawning in for-loops.

Figure~\ref{fig:for-example} shows a typical Go program where
several worker threads are concurrently sending data to the parent
thread via channel \texttt{a}.
Note that this program spawns $\rvert$\texttt{files}$\lvert$
threads and creates a channel whose capacity is
$\rvert$\texttt{files}$\lvert$.
The length of \texttt{files} is unknown at compile-time, hence this
program cannot be checked for concurrency errors with existing static
verification techniques for
Go~\cite{Ng15,Lange17,Lange18,Stadtmuller16,Midtgaard18}.

\smallpara{Our approach}
Our short-term objective is to improve the approach
from~\cite{Lange18} so that we can detect bugs in programs that
feature communication-related parameters that are unknown at
compile-time.
We focus on two kinds of communication-related parameters: ($i$) those
that determine the number of threads a program may spawn at run-time
and ($ii$) those that determine the capacity of channels.
For example, the number of threads and the capacity of channel
\texttt{a} are unknown at compile-time in
Figure~\ref{fig:for-example}. 
To fulfil our objective, we augment the behavioural types technique
of~\cite{Lange18} with
($i$) an \emph{intra}-procedural analysis to identify unknown
communication-related parameters, and ($ii$) a bounded verification
wrt.\ these parameters.
Concretely, we infer behavioural types from Go programs which may
feature (undefined) communication-related parameters.
If so, we ask the users to instantiate these parameters with bounds so
that we can model-check the inferred behavioural types.
The main challenges are to ask for user-provided bounds only when
necessary and to ensure that these bounds are used consistently.
We address these challenges by keeping track of variables that may be
used in channel creation statements or for-loops that spawn threads.

\begin{figure}
  \centering
  \includegraphics[width=0.8\textwidth]{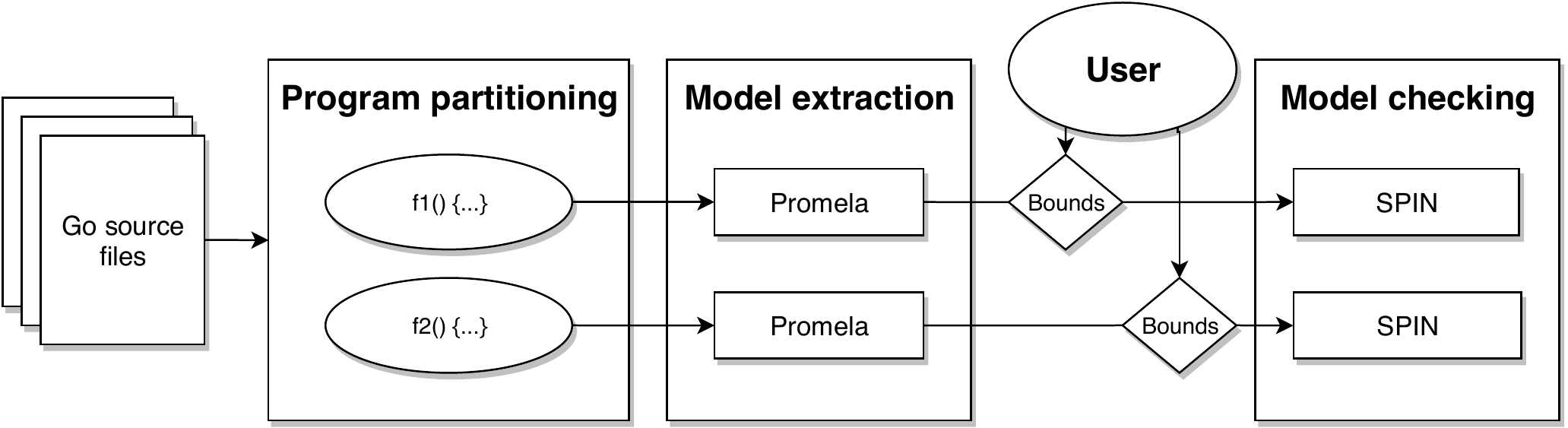}
  \caption{\togology\ workflow.} 
  \label{fig:process}
\end{figure}

Our long-term objective is to study automated \emph{repair} of
message-passing errors in Go.
To anticipate for this next step, we deviate from~\cite{Lange18} in
several ways.
(1) We infer behavioural types directly from Go source code instead of
its lower-level (SSA) representation.
(2) We use Promela and Spin instead of mCRL2 to encode and verify
behavioural types.
Promela has the advantage of being much closer to Go. It has an
imperative Go-like syntax and natively supports synchronous and
asynchronous channels.
As a consequence, it will be easier to syntactically map an
error in a Promela model to its source program.
(3) We divide Go programs into independent partitions. This allows us
to detect partial deadlocks and to identify the location of defects
more precisely, while making our tool faster.
Figure~\ref{fig:process} gives an overview of our approach, which we
have implemented in a tool called \gomela~\cite{gomela}.

\smallpara{Synopsis}
In \S~\ref{sec:minigo}, we present a core subset of Go, called
\minigo, as well as typical bugs that we want to rule out.
In \S~\ref{sec:promela-to-go}, we give a detailed algorithm to extract Promela
models from Go programs, while keeping track of communication-related
parameters.
In \S~\ref{sec:eval} we present our implementation and its empirical
evaluation.
We discuss related work and conclude in \S~\ref{sec:conc}.

\begin{figure}
\begin{lstlisting}[language=Go,basicstyle=\scriptsize\ttfamily,style=golang,firstline=3
]
func main() { /*<\label{line:main}>*/
	files := getFiles()                // decl. of getFiles() omitted 
	a := make(chan string, len(files)) // create bounded buffer 'a'/*<\label{line:new-chan}>*/
	for i := 0; i < len(files); i++ {  /*<\label{line:worker-spawn}>*/
		go worker(a, files[i], i) //spawn worker() /*<\label{line:spawned-for}>*/
	}
	for i := 0; i < len(files); i++ { /*<\label{line:for-loop}>*/
		<-a // receive from 'a' /*<\label{line:receive-a}>*/
	}
}
func worker(a chan int, f string, i int) {
	a <- parseFile(f, i) // send data on 'a' (decl. of parseFile() omitted)  /*<\label{line:send}>*/
}
\end{lstlisting}
\caption{File processing example}
\label{fig:for-example}
\end{figure}

\section{\minigo\ and message-passing concurrency errors} \label{sec:minigo}

For the sake of presentation, we use a fragment of Go that is
focused on its message-passing features and call it \minigo\ ---
we describe how our tool deals with a larger subset of Go in
Section~\ref{sec:eval}. The syntax of \minigo\ is given in
Figure~\ref{fig:syntax}.
We only discuss its semantics informally and refer to~\cite{Lange17}
for a formal account of the semantics of a variation of our language.

We use $\var$ to range over \emph{non-channel} variables, $\chan$ to
range over \emph{channel} variables, $\chanorvar$ to range over any
variables, $\expr$ to range over expressions (excluding channel
variables), $\chanexpr$ to range over expressions (possibly including
channel variables), $\id$ to range over function names, and $\num$ to
range over integer literals. We use $\forstmts$ to range over mutators
of for-loop indices.
We use $\args$ to range over a list of expressions and overload the
notation for
lists of statements ($\tilde{\stmt}$),  etc.
We write $\tilde\stmt_1 \tilde\stmt_2$ for the concatenation of
$\tilde\stmt_1$ and $\tilde\stmt_2$.
We write $\chansfunc(\args)$ (resp.\
$\chansfunc(\tilde{\chanorvar})$) for the maximal sub-list of $\args$
(resp.\ $\tilde{\chanorvar}$) that contains only channel variables.

A \minigo\ program $\prog$ consists of a list of function declarations
$\tilde\fundecl$, possibly including a \m{main} function (the
program's entry point).
Each function declaration specifies a list of parameters
$\tilde{\chanorvar}$ (possibly including channel variables) and a function
body $\stmts$.

Statement $\makechan$ creates a new channel of capacity
$\num$, when $\expr$ evaluates to $\num$.
If $\num =0$ then the channel is synchronous, otherwise it is
asynchronous.
Communication statements $\commstmt$ interact with channels:
$\var \leftarrow\textit{\chan}$ receives a value from channel $\chan$
and binds it to variable $\var$;
while $\chan \leftarrow \expr$ sends the evaluation of expression
$\expr$ on channel $\chan$.
Send actions are blocking when the channel is synchronous or has
reached its maximal capacity.
A channel can be closed with a $\close$ operation.
Any send or close action on a closed channel triggers a
run-time error.
Any receive action on a closed channel succeeds, if the channel is
empty a default value is returned.
Select statements $\select$ are guarded choices: they block until one
of the guarding communication operations succeeds; after which the
corresponding case is executed.
If multiple operations are available, one is chosen
non-deterministically.
Select statements may include a unique \m{default} branch, which is
taken if all other branches are blocking.

A statement $\gostmt$ spawns a new goroutine, i.e., an instance of
function $\id(\args)$ which is executed concurrently with its parent thread.
\minigo\ also includes standard constructs such as general
sequencing, conditionals, for-loops, and assignment.
For the sake of simplicity we model only the relevant parts of the
language of expression (see definition of $\expr$).
We assume that variable names are pairwise distinct. Additionally, as
in~\cite{Lange18}, we assume that channel are not in $\expr$, that
variables are immutable, and that recursive functions do not spawn
goroutines (for-loops are more common than recursion in Go).

\smallpara{Message-passing errors in \minigo}
In this work, we are interested in message-passing related bugs that \minigo\
programs may encounter at run-time. We distinguish three types of such bugs.
A \textbf{global deadlock} is a situation where at least one goroutine is
waiting for a send or receive action to succeed, while \emph{all} the other
goroutines are either blocked or terminated.
A \textbf{partial deadlock} is a situation where at least one goroutine is
permanently stuck while waiting for a send or receive action to succeed.
Go developers refer to partial deadlocks as \emph{goroutine leaks}
because such stuck goroutines never reach the end of their scope, and
thus are never garbage-collected.
A \textbf{channel safety error} is a situation where a send or close operation
is triggered on a closed channel.

\begin{figure}[t]
  \centering
$
\begin{array}{lcl}
\prog          & \coloneqq & \widetilde{\fundecl}\\[0.7mm]\stmt          & \coloneqq & \makechan\\
               &     |     & \commstmt \bnfor \select \bnfor  \close
  \\
               &     |     & \funcalldef \bnfor \gostmt \bnfor   \{ \stmts \} 
  \\
               &     |     & \ifelse  \bnfor \m{\text{for}} \ \var \texttt{:=} \expr_1 \m{;} \expr_2 \m{;}
                             \forstmts \, \{\stmts_3\}
  \\[0.7mm]
  \commstmt      & \coloneqq & \var \leftarrow\textit{\chan} \bnfor  \chan \leftarrow \expr
\end{array}
\quad
\begin{array}{lcl}
\commclause    & \coloneqq & \commclausedef \bnfor \default
  \\[0.7mm]
\expr          & \coloneqq & \true \bnfor \false \bnfor \num \bnfor \var \bnfor 
                               \ldots
  \\[0.7mm]
  \chanexpr      & \coloneqq & \chan \bnfor \expr
  \\[0.7mm]
  \chanorvar      & \coloneqq & \chan \bnfor \var
  \\[0.7mm]
  \fundecl       & \coloneqq & \fundecldef
  \\[0.7mm]
  \forstmts      & \coloneqq &             \var\texttt{++}
                        \mid
                        \var\texttt{--}
                        \mid
                        \ldots
\end{array}
$
\caption{Syntax of \minigo\ ($\chan$ ranges over channel variables and $\var$ stands for non-channel variables).}\label{fig:syntax}
\end{figure}

 \section{Extracting Promela models from \minigo\ programs} \label{sec:promela-to-go}
We adopt an approach based on behavioural types to produce a sound analysis of
\minigo\ for channel safety and global deadlock errors,
following~\cite{Lange18,Lange17}. 
Note that this approach is generally unsound wrt.\ liveness properties
such as partial deadlock freedom without a termination checker,
see~\cite[Section 5]{Lange17}.
In this context, behavioural types are an over-approximation of the
interactions between goroutines, i.e., they record send and receive
actions, while abstracting away from the computational aspects.
Typically, conditional statements are assigned behavioural types that correspond
to non-deterministic choices in process calculi.
In our work, behavioural types take the form of Promela models, which
we extract from \minigo\ source code.
Remarkably, we keep track of some computational aspects when they
affect the structure of the communication of the program, e.g., in
Figure~\ref{fig:for-example} we need to keep track of \texttt{len(files)}.

\begin{algorithm}[h]\scriptsize
  \SetAlgoLined \SetAlgoVlined \DontPrintSemicolon

  \Fn{\tt{\translateBlockStmt}$(\kbs,\stmt \stmts)$}{

    \Switch{$\stmt$}{
\lCase{$\chan \leftarrow \expr$}{
$\chan$\append\m{\promelacolor{.in!0;}}\append
        $\chan$\append\m{\promelacolor{.sending?state;}}\append
        \translateBlockStmt$(\kbs, \stmts)$ 
      }
      \lCase{$\var \leftarrow \chan$}{
        $\chan$\append\m{\promelacolor{.in?0;}}\append
        \translateBlockStmt$(\kbs, \stmts)$ 
      }
      \lCase{$\close$}{
        $\chan\append\m{\promelacolor{.closing?state;}} \append
        \translateBlockStmt(\kbs, \stmts)$ 
      }
      \Case{$\ifelse$}{
\m{\promelacolor{if}}\\
        \begin{tabular}{lll}
          &\m{\promelacolor{:: true ->}} &\append\translateBlockStmt$(\kbs, \stmts_1)$ \\
          &\m{\promelacolor{:: true ->}} &\append\translateBlockStmt$(\kbs, \stmts_2)$ \\
        \end{tabular}
        \\
        \m{\promelacolor{fi;}} \append\translateBlockStmt$(\kbs, \stmts)$ 
       
      }    
      \Case{$\select$}{
        \m{\promelacolor{if}} \\
        
          \begin{tabular}{lll}
          &\append\translateBlockStmt$(\kbs,\tilde \commclause)$ \\
          \end{tabular}\\
        \m{\promelacolor{fi;}} \append\translateBlockStmt$(\kbs, \stmts)$
      }
      \lCase{$\m{case } \commstmt : {\stmts_1}$}{
        \m{\promelacolor{:: }}\append\translateBlockStmt$(\kbs, \commstmt)$ \append \m{\promelacolor{-> }}\append\translateBlockStmt$(\kbs, \stmts_1)$
      }
      \lCase{$\m{default} : \{\stmts_1\}$}{
        \m{\promelacolor{:: true -> }}\append\translateBlockStmt$(\kbs, \stmts_1)$
      }
      \Case{$\funcalldef$}{
         \If{\tt{\texttt{\chansfunc}}$(\args)\neq []$}{

        \text{\m{\promelacolor{ch = [0] of \{int\}; run}}
          $\id\m{\promelacolor{(}}\texttt{\chansfunc} (\args),$ \m{\promelacolor{ch)}}\m{\promelacolor{;ch?0;}}}
        \append\translateBlockStmt$(\kbs, \stmts)$
        }\lElse{\translateBlockStmt$(\kbs, \stmts)$}       
      }
      \Case{$\m{go } \funcalldef$}{
        \If{\tt{\texttt{\chansfunc}}$(\args)\neq []$}{

          \m{\promelacolor{run go\_}}\append $\id$ \append \m{\promelacolor{(}}\texttt{\chansfunc}$(\args)$\m{\promelacolor{);\ }{\translateBlockStmt$(\kbs, \stmts)$}}
        } \lElse{\translateBlockStmt$(\kbs, \stmts)$}    
      }
      \Case{$\for$}{
        \texttt{\kwlet}\
        $(\kbs',x,y)$
\kweq\
\texttt{\mapcontains}$_\kbs(\var \texttt{:=} \expr \m{;} \expr \m{;} \forstmts)$ \kwin
\\
      \lIf{\tt{\spawns$(\stmts_3)\ \lor\       
          \kbs$ \colorize{\equals} $\kbs'$        
        }}{
          \\\m{{\promelacolor{for}}} {\promelacolor{(i: }}\append  \textit{x} \append\m{\promelacolor{..}}\append \textit{y} \append\m{\promelacolor{)\{}}\append\translateBlockStmt$(\kbs', \stmts_3)$\append\m{\promelacolor{\};}}
          \append\translateBlockStmt$(\kbs',\stmts)$
        }
        \lElse{

          \begin{tabular}{lll}
            \m{\promelacolor{do}}&\m{\promelacolor{:: true ->}}&\append\translateBlockStmt$(\kbs , \stmts_3)$\append\\
                                 &\m{\promelacolor{:: true ->}}&\m{\promelacolor{break;}}\\
            \m{\promelacolor{od;}}&
          \end{tabular}\\
          \append\translateBlockStmt$(\kbs, \stmts)$
        } 
      }
      \Case{$\makechan$}{
        \texttt{\kwlet}\
        $(\kbs',\_,y)$
        \kweq\
\texttt{\mapcontains}$_\kbs( i \coloneqq 0; \ i < \expr ;  \ i\texttt{++})$ \kwin  \\ 
        \m{\promelacolor{Chandef }}\append$\chan$\append \m{\promelacolor{;}}\\
        \m{\promelacolor{chan }}\append$\chan$\append\m{\promelacolor{.in = [}}\append$y$\append\m{\promelacolor{] of \{int\};}} \\
        \m{\promelacolor{run chanmonitor(}}\append$\chan$\append\m{\promelacolor{);}}\\
        \append\translateBlockStmt$(\kbs',\stmts)$
        
      }
}
    
  } 
  \caption{Extracting Promela from $\minigo$
    statements. We assume that $\translateBlockStmt(\kbs,[])$
    returns the empty string.}
  \label{algo:go-to-promela}
\end{algorithm}

Given a \minigo\ program $\prog$, we extract Promela models as follows.
For each function declaration $\fundecldef$ where $\tilde{\chanorvar}$ does
not contain any channel variables, we generate a model which
consists of three parts:
(1) a model entry point (\m{\promelacolor{init}} process in Promela)
that contains the translation of $\stmts$;
(2) a list of process declarations (\m{\promelacolor{proctype}} in Promela), one
for each distinct function call occurring (inter-procedurally) in
$\stmts$;
(3) a set of monitor processes, one for each channel created in $\stmts$.
Each of these models correspond to a partition of a \minigo\ program. Because
these partitions do not have free channel variables, they are effectively
independent.
Hence, we can verify them independently by considering each function declaration
without channel parameters as a program entry point.
As a consequence, we obtain a more precise and wider analysis of code-bases,
while reducing the computational cost of our analysis, comparing
to~\cite{Lange18}.
In particular, we can detect some \emph{partial} deadlocks in the
program under considering by identifying global deadlocks in some of its partitions.

Hereafter, we take the following conventions:
Promela strings generated by our algorithm are written in
\m{\promelacolor{typewriter blue}}. \minigo\ code is written in
$italic$.
Our approach is formalised through functions (in \texttt{typewriter
  black}) and algorithms (in {\bf{\colorize{bold-red}}}) that
manipulate \minigo\ programs.
Each identifier in \minigo\ is translated to the equivalent string in
Promela, e.g., $ch1$ in \minigo\ is translated to
\m{\promelacolor{ch1}}.
For the sake of readability, we omit the concatenation operator
between literal Promela strings and strings generated by translation
functions.

\smallpara{Function declarations}
Given a function body $\stmts$, for each distinct (blocking) function
call $\funcalldef$ occurring inter-procedurally in $\stmts$
such that $\chansfunc(\args)\neq []$, we define a Promela process
(\m{\promelacolor{proctype}}) as follows: 
\\
$
\m{\quad\qquad \promelacolor{proctype }}\append\id\append\m{\promelacolor{(}}\texttt{\chanParams}(\id)\m{\promelacolor{,ch)\{}}\append\translateBlockStmt(\varnothing, \body(\id)) \append\m{\promelacolor{;ch!0\}}}
$
\\
where \m{\promelacolor{ch}} is a channel used to signal the
termination of the function call (with \m{\promelacolor{ch!0\}}}).

\noindent
For each distinct non-blocking function call $\gostmt$,
such that $\chansfunc(\args)\neq []$,
we define the process:
\\ 
$
\m{\quad\qquad \promelacolor{proctype go\_}}\append\id\append\m{\promelacolor{(}}\texttt{\chanParams}(\id)
\m{\promelacolor{)\{}}\append\translateBlockStmt(\varnothing, \body(\id))\append \m{\promelacolor{\}}}
$

\noindent
where \texttt{\chanParams}$(\id)$ (resp.\ $\body(\id)$) returns the
\emph{channel} parameters (resp.\ body) of function $\id$ and $\varnothing$
denotes the empty map. 
Observe that the non-channel parameters are abstracted away.
We use \m{\promelacolor{proctype}} instead of
\m{\promelacolor{inline}} definition as the latter cannot include
declarations of new channels.
Next, we define function \translateBlockStmt\ which translates \minigo\ statements to Promela.

Algorithm~\ref{algo:go-to-promela} specifies how we extract a model from a list
of \minigo\ statements.
We use $\bound$ to range over the control statements
of a for-loop, i.e., $\bound$ ranges over triples of the form
($\var \texttt{:=} \expr_1 \m{;} \expr_2 \m{;} \forstmts$).

Function \translateBlockStmt\ takes two parameters: (1) $\kbs$ maps
expressions (corresponding to communication-related parameters) to
Promela strings, and (2) a list of $\minigo$ statements.

\smallpara{Channel primitives}
For each \minigo\ channel we generate a custom Promela structure,
called \m{\promelacolor{Chandef}}, which contains three channels:
\m{\promelacolor{in}} carries the exchanged messages, while
\m{\promelacolor{sending}} (resp.\ \m{\promelacolor{closing}}) is used
to monitor send (resp.\ closing) actions.
A send statement is translated to a send statement in Promela (on
channel \m{\promelacolor{in}}), followed by a receive statement on the
corresponding channel monitor (on the \m{\promelacolor{sending}}
channel, see below).
A receive statement is translated to a Promela receive (on channel
\m{\promelacolor{in}}).
A close statement is translated to a Promela receive on channel
\m{\promelacolor{closing}}.

\smallpara{Conditionals} 
An if-then-else statement is translated to an \m{\promelacolor{if}}
block in Promela. It behaves as a non-deterministic internal choice
(\m{\promelacolor{true}} is an always-enabled guard).
A select statement is translated to a non-deterministic choice, using
an \m{\promelacolor{if}} block where each non-default branch is
guarded by a send or receive action.
Default branches are translated to a branch that is always available.

\begin{figure}[t]
\footnotesize
  \SetAlgoLined \SetAlgoVlined \DontPrintSemicolon 

\[
  \lowerBound{b} =
  \begin{cases}
    e_1
    &
    \text{if } b \text{ is } \var \coloneqq \expr_1; \ \var < \expr_2 ;  \ \var\texttt{++}
    \\
    e_2
    &
    \text{if } b \text{ is } \var \coloneqq \expr_1; \ \var > \expr_2 ;  \ \var\texttt{--}
    \\
    \bot & \text{otherwise}
  \end{cases}
  \qquad  \qquad
  \upperBound{b} =
  \begin{cases}
    e_2
    &
    \text{if } b \text{ is } \var \coloneqq \expr_1; \ \var < \expr_2 ;  \ \var\texttt{++}
    \\
    e_1
    &
    \text{if } b \text{ is } \var \coloneqq \expr_1; \ \var > \expr_2 ;  \ \var\texttt{--}
    \\
    \bot & \text{otherwise}
  \end{cases}
\]
\[
\text{\tt\mapcontains}_\kbs(\bound) =
\begin{cases}
    (\kbs, {{{x}}}, {{{y}}})
    & \text{if }  \lowerBound(\bound) = \bot \text{ or } \upperBound(\bound) = \bot,
    \text{ with } {{{x}}} \text{ and } {{{y}}}  \text{ fresh}
    \\
  ( \kbs',
  \kbs'(\lowerBound{\bound}) ,\kbs'(\upperBound{\bound})
  )
  &
  \text{otherwise, where }
  \kbs' = \kbs \left[
    \expr \mapsto {{{x}}}
    \, \mid \,
    \expr  \in
    \left\{  \lowerBound(\bound), \upperBound(\bound)  \right\}  \setminus \domain(\kbs) \text{ with } {{{x}}} \text{ fresh}
  \right]
  \end{cases}
\]
\caption{Auxiliary functions for Algorithm~\ref{algo:go-to-promela},
  where we assume $\kbs(\num) = \num$ for each integer $\num$.}
  \label{fig:lookup}
\end{figure}
 
\smallpara{Control statements}
A blocking function call is translated to Promela code that spawns an
instantiation of the corresponding Promela process (using the
\m{\promelacolor{run}} keyword), then waits for it to terminate by
waiting on fresh channel \m{\promelacolor{ch}}.
For spawning function calls, i.e., $\m{go } \funcalldef$, there are two cases.
If the parameters include channels, the algorithm returns Promela code
that spawns the corresponding process.
Otherwise it omits the call entirely --- as it will be checked in the
model of an independent partition.

To translate $\for$, we need to first check ($i$) whether we can
extract well-identified bounds that we consider as
communication-related parameters and ($ii$) whether the loop contains
(inter-procedurally) spawning function calls, i.e.,
\texttt{\spawns$(\stmts_3)$} holds (whose straightforward definition
is omitted).
If \texttt{\spawns$(\stmts_3)$} holds then the range of the loop needs
to be finite. Hence, either we set-up Promela variables (that the user
will instantiate) to define a range; or we are able to identify a
static range (integer literal bounds).
When the loop does not spawn new threads, we only use a finite Promela
for-loop if all involved variables have been seen before
($\kbs \equals \kbs'$, see below).
In all other cases, we use a non-deterministic loop using a Promela
\m{\promelacolor{do}} block which can be exited at any iteration with
a \m{\promelacolor{break}} operation.

We keep track of re-usable communication-related parameters with the
map $\kbs$ and 
use the function \mapcontains, defined in Figure~\ref{fig:lookup},
to query it.
Functions $\lowerBound(\bound)$ and $\upperBound(\bound)$ respectively
return the lower and upper bounds of for-loop control statement $\bound$.
When the control statements of a for-loop are well-formed (they obey a
recognisable pattern, i.e., $ \lowerBound{\bound} \neq \bot \land
\upperBound{\bound}\neq\bot$), the \mapcontains\ function returns a new map
$\kbs'$ and the lower ( $\kbs'(\lowerBound{\bound})$) and upper
($\kbs'(\upperBound{\bound})$) bounds of the for-loop.
Map $\kbs'$ augments $\kbs$ with mappings from newly identified
\emph{expressions} to fresh Promela variables.

\smallpara{Channel creation}
A channel creation statement is translated to the instantiation of a
\m{\promelacolor{Chandef}} structure and the spawning of its
\m{\promelacolor{chanmonitor}} process.
We initialise the \m{\promelacolor{in}} channel of the
\m{\promelacolor{Chandef}} structure with a capacity corresponding to
its \minigo\ equivalent.
If the capacity is not a integer literal, the \mapcontains\ function ensures
that we either re-use Promela variables, or generate fresh ones.
Note that channels \m{\promelacolor{sending}} and
\m{\promelacolor{closing}} in \m{\promelacolor{Chandef}} are always
synchronous.

\begin{wrapfigure}{r}{0.4\textwidth}
\begin{tikzpicture}[ initial text={}, node distance=0.8cm and 1.6cm, ->,>=stealth,font=\scriptsize,
    scale=0.9, every node/.style={transform shape}
    ,initial distance=0.25cm]
    \node[state, initial, initial where=above,accepting] (open) {open};
    \node[state,right=of open,accepting] (close) {closed};
    \node[state,right=of close,fill=red!25] (error) {error};
\path
    (open) edge [loop below] node  {\m{\promelacolor{ch.sending}}!} (open)
    (open) edge  node  [above] {\m{\promelacolor{ch.closing}}!} (close)
    (close) edge [bend right] node  [below] {\m{\promelacolor{ch.closing}}!} (error)
    (close) edge [bend left] node  [above] {\m{\promelacolor{ch.sending}}!} (error)
    (close) edge [loop above] node  {\m{\promelacolor{ch.in}}!} (close)
    ;
  \end{tikzpicture}
\end{wrapfigure}

 \smallpara{Channel monitors}
To detect channel safety errors, we keep track of the state of \minigo\
channels. We use Promela processes to monitor channel actions, i.e., send,
receive, and close.
As an optimisation, we only create such monitors when a \texttt{close} primitive
appears in the program.
Processes corresponding to goroutines interact with channel monitors via an
instance \m{\promelacolor{ch}} of a \m{\promelacolor{ChanDef}} structure which
contains three channels (\m{\promelacolor{in}}, \m{\promelacolor{sending}}, and
\m{\promelacolor{closing}}).
The automaton on the right represents the behaviour of this monitor
(\closedChan). Figure~\ref{fig:for-example-pml}
(lines~\ref{line:chan-monitor}-~\ref{line:chan-monitor-end}) gives the
corresponding Promela code.
In \minigo\ and Go, when a channel is \textit{closed}, sending on it or
closing it will raise an error, hence the transitions from state
\textit{closed} to state \textit{error} in
the automaton.
Also, receive actions on closed channels always succeed, hence the self-loop on
state \textit{closed}.

 \begin{figure}
\begin{lstlisting}[basicstyle=\scriptsize\ttfamily,language=promela,firstline=0,multicols=2]
#define len_files_0  15 /*<\label{line:global-var}>*/

typedef Chandef {
  chan in = [0] of {int};
  chan sending = [0] of {int};
  chan closing = [0] of {bool};
}
init {  /*<\label{line:init}>*/
  chan _ch0_in = [len_files_0] of {int};
  Chandef _ch0;
  _ch0.in = _ch0_in;
  run chanMonitor(_ch0) /*<\label{line:spawn-chan-monitor}>*/

  for(i : 1.. len_files_0) {
    run mainworker(_ch0)
  };
  for(i : 1.. len_files_0) {
    _ch0.in?0
  };
} /*<\label{line:init-end}>*/
proctype mainworker(Chandef a) { /*<\label{line:worker}>*/
  a.in!0;
  a.sending?0
} /*<\label{line:worker-end}>*/

proctype chanMonitor(Chandef ch) { /*<\label{line:chan-monitor}>*/
  bool open = true;
  do /*<\label{line:do}>*/
  :: true -> 
    if
    :: open -> 
      if /*<\label{line:end1}>*/
      :: ch.sending!open; /*<\label{line:send_on_open}>*/
      :: ch.closing!true -> /*<\label{line:closing1}>*/
        open = false
      fi
    :: else -> 
      if /*<\label{line:if_server}>*/
      :: ch.sending!open ->  /*<\label{line:send_on_close}>*/
        assert(false) /*<\label{line:assert_false}>*/
      :: ch.closing!true -> /*<\label{line:closing}>*/
        assert(false) /*<\label{line:assert_false1}>*/
      :: ch.in!0; /*<\label{line:send_in}>*/
      fi
    fi
  od
}/*<\label{line:chan-monitor-end}>*/
\end{lstlisting}
  \caption{Model extracted from Listing~\ref{fig:for-example} with Algorithm~\ref{algo:go-to-promela}}
  \label{fig:for-example-pml} 
\end{figure}
 \begin{example}
  The model extracted from Figure~\ref{fig:for-example} with
  Algorithm~\ref{algo:go-to-promela} is given in
  Figure~\ref{fig:for-example-pml}.
Lines \ref{line:init}-\ref{line:init-end} contain the translation of the
  \texttt{main} function.
Note that a \closedChan\ is spawned at
  line~\ref{line:spawn-chan-monitor}. The definition of this process
  is given in
  lines~\ref{line:chan-monitor}-\ref{line:chan-monitor-end}.
The translation of the \texttt{worker} function is
  in lines~\ref{line:worker}-\ref{line:worker-end}.
  
  Figure~\ref{fig:for-example-pml} contains one (unknown) communication-related
  parameter: \texttt{len(files)} which is used in the capacity of channel
  \texttt{a}, and in the loops at lines~\ref{line:worker-spawn}
  and~\ref{line:for-loop} of Figure~\ref{fig:for-example}.
Observe that the communication-related parameter \texttt{len(files)}
  is set to 15 here (see line \ref{line:global-var}). Our
  implementation also allows user to specify such parameters as program
  (command line) arguments.
Expression \texttt{len(files)} is first seen by
  Algorithm~\ref{algo:go-to-promela} in a channel creation statement.
At this point, it adds \texttt{len(files)} $\mapsto$
  \texttt{`len\_files\_0'} in map $\kbs$.
When the algorithm processes both loops, it invokes
  \mapcontains$_\kbs(\texttt{i:=0;i<len(files);i++})$ to obtain lower
  and upper bounds ($0$ and \texttt{`len\_files\_0'}, respectively).
Note that the loop in line~\ref{line:worker-spawn} is a spawning loop,
  hence it must be translated to a finite loop in Promela to obtain a finite
  model.
The loop in line~\ref{line:for-loop} is not spawning, but it is still
  translated to a finite loop because all its bounds are already in $\kbs$.
\end{example}

\section{Implementation and evaluation}\label{sec:eval}

\newcounter{ExampleCounter}  
\newcounter{ColumnCounter}
\renewcommand{\theColumnCounter}{\arabic{ColumnCounter}}
\newcommand{\evfunname}[1]{{\tt\scriptsize #1}}

\begin{table}[t]
  \caption{\small Go programs verified by \togology\ and comparison
    with Godel~\cite{Lange18}.  All programs are available online~\cite{gomela}.  } \label{tbl:applicability}
  \begin{center}
    \renewcommand{\arraystretch}{0.75}
    {\scriptsize
      \renewcommand{\tabcolsep}{0.12cm}
      \begin{tabular}{r || l || r | c | c | r  r || r | c | c || c | c  r }
        \toprule
        &&
            \multicolumn{5}{c||}{\textbf{\togology}} 
        & \multicolumn{3}{c||}{\textbf{Godel}}
        &  \multicolumn{2}{c}{\textbf{Manual analysis}}
        \\
        \# & \textbf{Programs \& \texttt{Partitions}} & \textbf{$\lvert$States$\rvert$}  & \textbf{CS} & \textbf{GD} & \textbf{Infer} \tiny{(ms)} & \textbf{Spin} \tiny{(ms)} & \textbf{Time} & $\psi_s$ & $\psi_g$ & \textbf{\; \, CS\; \, } & \textbf{GD} \\
        \midrule 
        \refstepcounter{ExampleCounter}\arabic{ExampleCounter}\label{ex:runningex}&
                                                                file processing ($\rvert$\texttt{files}$\lvert$=15) & 376880 & \cmark & \cmark & 85 & 1666 & $\bot$ & $\bot$ & $\bot$ & \cmark & \cmark\\
        \refstepcounter{ExampleCounter}\arabic{ExampleCounter}\label{ex:runningex-gd}&
                                                                file processing v1 ($\rvert$\texttt{files}$\lvert$=15) & 109 & \cmark & \xmark & 86 & 1057 & $\bot$ & $\bot$ & $\bot$ & \cmark & \xmark\\
        \refstepcounter{ExampleCounter}\arabic{ExampleCounter}\label{ex:runningex-leak}&
                                                                file processing v2 ($\rvert$\texttt{files}$\lvert$=15) & 110 & \cmark & \xmark & 85 & 1027 & $\bot$ & $\bot$ & $\bot$ & \cmark & \xmark\\
        \refstepcounter{ExampleCounter}\arabic{ExampleCounter}\label{ex:prodcons}&
                                                                prod-cons (\texttt{k}=5,\texttt{n}=10,\texttt{m}=10) & 4802785 & \cmark & \cmark & 85 & 31239 & $\bot$ & $\bot$ & $\bot$ & \cmark & \cmark\\
         \refstepcounter{ExampleCounter}\arabic{ExampleCounter}\label{ex:altbit}&
                                                                alt-bit & 35 & \cmark & \cmark & 956 & 1391 & 460& \cmark & \cmark & \cmark & \cmark\\
        \refstepcounter{ExampleCounter}\arabic{ExampleCounter}\label{ex:concsys}&
                                                                concsys & 96 & - & - & 768 & 8194 & 588& \cmark & \cmark & - & - \\
        & - \evfunname{ConcurrentSearch()} & 49 & \cmark & \cmark &  -  & 1126 &  - & - & - & \cmark & \cmark \\
        & - \evfunname{ConcurrentSearchWC()}& 26 & \cmark & \xmark &  -  & 1335 &  - & - & - & \cmark & \xmark \\
        & - \evfunname{First()} & 3 & \cmark & \cmark &  -  & 1183 &  - & - & - & \cmark & \cmark \\
        & - \evfunname{ReplicaSearch()} & 9 & \cmark & \xmark &  -  & 1326 &  - & - & - & \cmark & \xmark \\
        & - \evfunname{SequentialSearch()}& 3 & \cmark & \cmark &  -  & 1073 &  - & - & - & \cmark & \cmark \\
        & - \evfunname{FakeSearch()}& 3 & \cmark & \cmark &  -  & 1083 &  - & - & - & \cmark & \cmark \\
        & - \evfunname{main()}& 3 & \cmark & \cmark &  -  & 1068 &  - & - & - & \cmark & \cmark \\
        \refstepcounter{ExampleCounter}\arabic{ExampleCounter}&
                                                                cond-recur & 13 & \cmark & \cmark & 84 & 1278 & 623& \cmark & \cmark & \cmark & \cmark\\
        \refstepcounter{ExampleCounter}\arabic{ExampleCounter}&
                                                                dinephil & 968 & \cmark & \cmark & 804 & 1478 & 859& \cmark & \cmark & \cmark & \cmark\\
        \refstepcounter{ExampleCounter}\arabic{ExampleCounter}\label{ex:dinephil5}&
                                                                dinephil5 & 71439 & \cmark & \cmark & 835 & 1801 & 8681& \cmark & \cmark & \cmark & \cmark\\
        \refstepcounter{ExampleCounter}\arabic{ExampleCounter}\label{ex:double-close}&
                                                                double-close & 18 & \xmark & - & 85 & 1486 & 463& \xmark & \cmark & \xmark & \cmark\\
        \refstepcounter{ExampleCounter}\arabic{ExampleCounter}\label{ex:fanin-alt}&
                                                                fanin-alt & 34 & \cmark & \xmark & 769 & 1686 & 786& \cmark & \cmark & \cmark & \xmark\\
        \refstepcounter{ExampleCounter}\arabic{ExampleCounter}\label{ex:fanin}&
                                                                fanin & 15 & \cmark & \cmark & 681 & 1495 & 621& \cmark & \cmark & \cmark & \cmark\\
        \refstepcounter{ExampleCounter}\arabic{ExampleCounter}&
                                                                fixed & 14 & - & - & 740 & 2379 & 656& \cmark & \cmark & - & - \\
        & - \evfunname{Work()}& 2 & \cmark & \cmark & - & 1117 & - & - & - & \cmark & \cmark\\
        & - \evfunname{main()}& 12 & \cmark & \cmark & - & 1262 & - & - & - & \cmark & \cmark\\
        \refstepcounter{ExampleCounter}\arabic{ExampleCounter}&
                                                                forselect & 21 & \cmark & \cmark & 755 & 1507 & 813& \cmark & \cmark & \cmark & \cmark\\
        \refstepcounter{ExampleCounter}\arabic{ExampleCounter}&
                                                                jobsched & 48 & - & - & 781 & 2745 & 589& \cmark & \cmark & - & - \\
        & - \evfunname{main()} & 45 & \cmark & \cmark & - & 1532 & - & - & - & \cmark & \cmark \\
        & - \evfunname{morejob()} & 3 & \cmark & \cmark & - & 1213 & - & - & - & \cmark  & \cmark \\
        \refstepcounter{ExampleCounter}\arabic{ExampleCounter}\label{ex:mismatch}&
                                                                mismatch & 12 & - & - & 714 & 2316 & 603& \cmark & \xmark & - & - \\
        &  - \evfunname{Work()}& 2 & \cmark & \cmark & - & 1044 & - & - & - & \cmark & \cmark\\
        &  - \evfunname{main()}& 10 & \cmark & \xmark & - & 1272 & - & - & - & \cmark & \xmark\\
        \refstepcounter{ExampleCounter}\arabic{ExampleCounter}\label{ex:sel}&
                                                                sel & 21 & \cmark & \xmark & 84 & 1719 & 326 & \cmark & \xmark & \cmark & \xmark\\
        \refstepcounter{ExampleCounter}\arabic{ExampleCounter}&
                                                                selFixed & 19 & \cmark & \cmark & 82 & 1330 & 572& \cmark & \cmark & \cmark & \cmark\\
        \refstepcounter{ExampleCounter}\arabic{ExampleCounter}\label{ex:philo}&
                                                                philo & 18 & \cmark & \xmark & 85 & 1362 & 537& \cmark & \xmark & \cmark & \xmark\\
        \refstepcounter{ExampleCounter}\arabic{ExampleCounter}\label{ex:starvephil}&
                                                                starvephil & 67 & \cmark & \xmark & 759 & 1343 & 836& \cmark & \xmark & \cmark & \xmark\\
        \refstepcounter{ExampleCounter}\arabic{ExampleCounter}\label{ex:non-live}&
                                                                nonlive & 7 & \cmark & \cmark & 850 & 1194 & 550& \cmark & \cmark & \cmark & \cmark\\
        \refstepcounter{ExampleCounter}\arabic{ExampleCounter}\label{ex:non-live_v1} &
                                                                nonlive v1 & 7 & \cmark & \xmark$^\dagger$ & 850 & 1152 & 366 & \cmark & \xmark$^\dagger$ & \cmark & \cmark\\
        \refstepcounter{ExampleCounter}\arabic{ExampleCounter}&
                                                                prod-cons & 61 & \cmark & \cmark & 87 & 1390 & 508& \cmark & \cmark & \cmark & \cmark\\
        \refstepcounter{ExampleCounter}\arabic{ExampleCounter}\label{ex:prodcons3}&
                                                                prod3-cons3 & 5746 & \cmark & \cmark & 643 & 1534 & 19963& \cmark & \cmark & \cmark & \cmark\\
        \refstepcounter{ExampleCounter}\arabic{ExampleCounter}\label{ex:prodconsclose}&
                                                                prodconsclose & 12185424 & \cmark & \cmark & 163 & 18672 & 25348& \cmark & \cmark & \cmark & \cmark\\
        \refstepcounter{ExampleCounter}\arabic{ExampleCounter}\label{ex:stuckmsg}&
                                                                stuckmsg & 5 & \cmark & \cmark & 86 & 1109 & 790& \cmark & \cmark & \cmark & \cmark\\
        \refstepcounter{ExampleCounter}\arabic{ExampleCounter}\label{ex:data-dependent}&
                                                                data-dependent & 12 & \cmark & \xmark$^\dagger$ & 86 & 1170 & 694 & \cmark & \xmark$^\dagger$ & \cmark & \cmark\\
        \midrule
        &Column number & 1 & 2 & 3 & 4 & 5 & 6 & 7 & 8 & 9 & 10
        \\
        \bottomrule
      \end{tabular}
    }
  \end{center}
\end{table}

\smallpara{Implementation}
We have implemented our approach in a tool called
\gomela~\cite{gomela}.
Given a Go program, our tool extracts Promela models (as described in
Section~\ref{sec:promela-to-go}).
If necessary, the user enters values (bounds) for the statically unknown
communication-related parameters produced by Algorithm~\ref{algo:go-to-promela}
(i.e., the Promela variables).
For instance, the user provides a bound to instantiate \texttt{len(files)} in
Figure~\ref{fig:for-example}. This value is then used as a bound in the
for-loops at lines~\ref{line:worker-spawn} and~\ref{line:for-loop}
as well as the capacity of channel \texttt{a}.

\gomela\ uses Spin to check whether each model is free from channel
safety errors and global deadlocks.
Spin reports any global deadlock and any trace that leads to an
\m{\promelacolor{assert(false)}} statements. We use the former to
check for global deadlocks (GD) in a program's partitions, and the
latter to check for channel safety errors (CS).
Spin detects when the main process (\m{\promelacolor{init}})
terminates while another process is still running. We use this to
detect some goroutine leaks, i.e., a particular case of partial
deadlocks.

In addition to \minigo\ statements, \gomela\ deals with constants
(used as communication-related parameters), anonymous functions, break
statements, for range loops and switch statements.
Occurrences of integer constants are replaced with their actual
values.
To deal with anonymous functions, \gomela\ generates Promela
corresponding function declarations (using fresh names) and its
(unique) invocation.
Go's \texttt{break} statements are translated as Promela break
statements.
For-ranges loop (\m{for range} $list \{ \stmts_1\}$) are treated as
for-loops with control statements of the form: to
$\m{\text{for}} \ i\texttt{:=0} \m{;} i < len(list) \m{;}
i\texttt{++}$.
Finally, \texttt{switch} statements are translated into n-ary internal
choices (similar to an if-then-else).

\smallpara{Evaluation} 
To evaluate our tool, we ran it on several benchmarks, including some
from~\cite[Table 1]{Lange18}.
The results of this evaluation are in Table~\ref{tbl:applicability}.
In Table~\ref{tbl:applicability}, Column~1 gives the number of states in the
model, as given by Spin.
In Column~2 (resp.~3) a \cmark-mark says that the corresponding program
partition is channel-safe (resp.\ free of global deadlock); a
\xmark-mark says that the property is violated.
Column~4 (resp.~5) shows the time (milliseconds) taken to extract (resp.\
verify) the Promela model of a partition. The timing for programs are the sum of
all of their partitions.
Column~6 shows the time (milliseconds) taken by Godel~\cite{Lange18} to verify
channel safety ($\psi_s$, Column~7) and global deadlock ($\psi_g$, Column~8)
properties in.
A $\bot$-mark means that Godel does not support this program.
A \cmark-mark in Column~9 (resp.~10) says that it was not possible to
find any channel errors (resp.\ global deadlocks) manually,
\xmark$^\dagger$ highlights false alarms.

In Table~\ref{tbl:applicability}, Program~\ref{ex:runningex} is the
example in Figure~\ref{fig:for-example} where the user has set
\texttt{len(files)} to 15.
Programs~\ref{ex:runningex-gd} and~\ref{ex:runningex-leak} are variations of
Program~\ref{ex:runningex} with a global deadlock and a goroutine leak,
respectively.  
Program~\ref{ex:prodcons} spawns \texttt{n} producers and \texttt{m} consumers
that interact (repeatedly) over a channel with capacity \texttt{k}.
Observe that these are not supported by Godel because they include
thread-spawning in a for-loop.
Programs~\ref{ex:altbit} to~\ref{ex:stuckmsg} are taken
from~\cite[Table 1]{Lange18}.
We note that Godel runs marginally faster than \togology\ on programs
with small models.
This is due to Spin's start up time of $\sim$1s.
For larger programs (more than 5000 states) \togology\ performs much
better than Godel, see, e.g., Programs~\ref{ex:dinephil5},
\ref{ex:prodcons3}, and~\ref{ex:prodconsclose}.
This suggests that our tool scales better than Godel on larger
code-bases.

Below we comment on the errors identified in the programs from
Table~\ref{tbl:applicability}.
\begin{itemize}[leftmargin=2.5em, labelwidth=!, labelindent=0pt]\itemsep-2pt
\item Program~\ref{ex:concsys} has two partitions that contain global
deadlocks.
In \texttt{ConcurrentSearchWC}, a function spawns three goroutines
that send a message on a shared channel \texttt{c}. That function may
terminate silently (via timeout) hence leaving three unmatched send actions
in the goroutines.
\texttt{ReplicaSearch} includes a similar pattern where the parent
thread may terminate while leaving some goroutines permanently blocked.
\item Program~\ref{ex:double-close} contains a channel safety error where a
  channel is closed twice by two different threads. We note that
  Spin aborts the verification as soon as it finds such errors.
\item Program~\ref{ex:fanin-alt} creates two producers and one consumer. The
  consumer may terminate silently in which case both producers are
  blocked permanently.
Program~\ref{ex:fanin} is similar to Program~\ref{ex:fanin-alt} except
  that the consumer never terminates, thus fixing the bug.
\item Program~\ref{ex:mismatch} contains two partitions: \texttt{Work} 
  consist of a non-terminating loop (without communications), while
\texttt{main} contains a global deadlock due to a mismatch between the
  number of send and receive actions.
Note that if we were modelling this program with one monolithic
  partition, we would not detect a global deadlock because
  \texttt{Work} never blocks.
\item In Program~\ref{ex:sel} each goroutine may get stuck because of a
  mismatch between the number of send and receive actions.
\item Program~\ref{ex:philo} is an encoding of the (starving)
  philosopher problem using only two philosophers, taken
  from~\cite{Stadtmuller16}.
Program~\ref{ex:starvephil} is another encoding of the (starving)
  dining philosophers from~\cite{Lange18}.
\item Programs~\ref{ex:non-live} and~\ref{ex:non-live_v1} are two
  variants of a program that contain two goroutines: one is waiting
  for a message, while the other contains a non-terminating for-loop.
In Program~\ref{ex:non-live}, the non-terminating for-loop is
  followed by a (unreachable) matching send action.
These programs show the limits of the behavioural types approach:
  they contain proper partial deadlocks.
The problem in Programs~\ref{ex:non-live} is only detected in Godel
  by using an additional termination checker.
\item Program~\ref{ex:data-dependent} is given in
  Figure~\ref{fig:data-dependent}.
This program gives a typical example of a false alarm raised by our
  tool, and any existing approach based on behavioural types.
Because if-then-else statements are translated to
  non-deterministic choices, our approach is unable to determine that
  the two conditional blocks are ``synchronised'' by the same invocation of
  function \texttt{f()}.  
\end{itemize}

\begin{figure}
\begin{lstlisting}[language=Go,basicstyle=\footnotesize\ttfamily,style=golang,multicols=2,firstline=2
]
func main() {
	a := make(chan int)
	go send(a)
	if f() { /*<\label{line:if1}>*/ // decl. of f() elided
		a <- 0
	} else {}
}

func send(a chan int) {
	if f() { /*<\label{line:if2}>*/
		<-a
	} else {}
}
/* f() is a deterministic function without side-effects */
\end{lstlisting}
\caption{Data-dependent choice (Program~\ref{ex:data-dependent}).}\label{fig:data-dependent}
\end{figure}
 \smallpara{Limitations}
Our approach is applicable to \minigo, extended with the syntactic
constructs discussed above. Several key limitations need to be tackled
to address the full Go language.
We assume that variables are immutable, as a consequence we cannot
soundly analyse programs that, e.g., mutate a list \texttt{files} in
between using \texttt{len(files)} as a communication-related
parameter.
Go has object oriented-like features, such as structs, methods, and
interfaces which we currently do not support. Virtual method calls (on
interfaces) are particularly difficult to model.
As in~\cite{Lange18, Lange17,Ng15,Stadtmuller16,Midtgaard18}, we do
not support channel passing (since we abstract away the data sent over
channels).
We note that our empirical survey~\cite{NDilley19} found that only 6\%
of projects used channels that carry channels.

 \section{Related work, conclusions and future work}\label{sec:conc}

\smallpara{Related work}
Spin and Promela have been used extensively in software
verification.
Notably Java PathFinder~\cite{Havelund00} translates Java
programs to Promela models which are then verified for deadlocks and
violations of user-provided assertions.
Also, Zaks and Joshi~\cite{Zaks08} use Spin to verify
multi-threaded C programs using their LLVM representation and custom
virtual machine.

Several works focus on the verification of message-passing concurrency in
Go~\cite{Ng15,Lange17,Lange18,Stadtmuller16,Midtgaard18,Sulzmann17,Sulzmann18}.
Four papers studied static verification using behavioural models.
Ng and Yoshida~\cite{Ng15} proposed $dingo\-hunter$, the first static
global deadlock detection tool for Go. It relies on communicating
finite-states machines~\citep{Brand83} and multiparty
compatibility~\cite{Lange15}.
Their work does not support asynchronous channels nor programs that
spawn goroutines or create channels in loops or conditionals.
Stadtm{\"u}ller et al.\ introduced $gopherlyzer$~\cite{Stadtmuller16}
which detects global deadlocks using forkable regular expressions.
This work does not support channel closures, asynchronous channels,
nor goroutines spawned in loops.
Lange et al.~\cite{Lange17,Lange18} proposed Gong and Godel, whose
approach serves as a basis for this work.
Gong uses an \emph{ad-hoc} checker which supports bounded
verification of infinite-state models, but did not scale well.
Instead, Godel uses mCRL2~\cite{CranenGKSVWW13} as a back-end. Because
mCRL2's communication model is very different from Go's the encoding
from behavioural types to mCRL2's language is very
intricate, see~\cite{godelgit}.
Promela is a more convenient language for this purpose, but because
Spin supports only LTL, while mCRL2 supports the $\mu$-calculus, it is
not possible to check the liveness property specified
in~\cite{Lange17,Lange18}.
However, we can still identify some goroutine leaks by checking
whether their corresponding models reach their end states.

\smallpara{Conclusions}
Our work builds on the approach in~\cite{Lange18} and improves it to
support statically unknown communication-related parameters via a
bounded analysis.
Our approach allows us to support programs that spawn a parameterised
number of goroutines or channel capacities.
Our evaluation shows that our tool scales well and produces models
that can be easily understood and adjusted by programmers.

\smallpara{Future work}
Our short term plans are to support additional concurrency-related Go
features, e.g., barriers (\texttt{WaitGroup}).
We will also improve our algorithms to support more complex for-loops
control statements, and to perform a fully \emph{inter}-procedural
analysis of communication-related parameters.
In the longer term, we plan to use our tool to detect concurrency
errors and suggest repairs for large code-bases.
We plan to perform a large-scale empirical evaluation of this
toolchain on the dataset identified in~\cite{NDilley19}.

\bibliographystyle{eptcs}
\bibliography{bib}

\end{document}